\documentclass[aps,prl,twocolumn,showpacs,amsmath,amssymb,superscriptaddress]{revtex4-1}
\usepackage[utf8]{inputenc}

\usepackage{tikz}
\usepackage{graphicx}
\usepackage{amsfonts}
\usepackage{amsmath}
\usepackage{amssymb,bbold}
\usepackage{color}
\usepackage{bm}
\usepackage{bbm}
\usepackage{enumerate}
\usepackage{amsfonts, amsmath, amsthm, amssymb} 
\usepackage{mathtools}
\usepackage{array}
\usepackage[colorlinks,bookmarks=false,citecolor=blue,linkcolor=red,urlcolor=blue]{hyperref}

\begin{document}

\title{Anomalous diffusion and localization in a positionally disordered quantum spin array}

\author{Rapha\"el Menu}
\affiliation{Univ Lyon, Ens de Lyon, Univ Claude Bernard, CNRS, Laboratoire de Physique, F-69342 Lyon, France}
\author{Tommaso Roscilde}
\affiliation{Univ Lyon, Ens de Lyon, Univ Claude Bernard, CNRS, Laboratoire de Physique, F-69342 Lyon, France}
\date{\today}

\begin{abstract}
 Disorder in quantum systems can lead to the disruption of long-range order in the ground state and to the localization of the elementary excitations - famous examples thereof being the Bose glass of interacting bosons in a disordered or quasi-periodic environment, or the localized phase of spin chains mapping onto fermions. Here we present a two-dimensional quantum Ising model - relevant to the physics of Rydberg-atom arrays - in which positional disorder of the spins induces a randomization of the spin-spin couplings and of an on-site longitudinal field. This form of disorder preserves long-range order in the ground state, while it localizes the elementary excitations above it, faithfully described as spin waves: the spin-wave spectrum is partially localized for weak disorder (seemingly exhibiting mobility edges between localized and extended, yet non-ergodic states), while it is fully localized for strong disorder. The regime of partially localized excitations exhibits a very rich non-equilibrium dynamics following a low-energy quench: correlations and entanglement spread with a power-law behavior whose exponent is a continuous function of disorder, interpolating between ballistic and arrested transport. Our findings expose a stark dichotomy between static and dynamical properties of disordered quantum spin systems, which is readily accessible to experimental verification using quantum simulators of closed quantum many-body systems.         
\end{abstract}

\maketitle

 Anderson localization of waves in a disordered environment is a striking phenomenon fundamentally relying on the coherent nature of wave dynamics \cite{Anderson-book,EversM2008}. When cast in the context of quantum mechanics, Anderson localization can manifest itself either at the single particle level, as localization of wavefunctions (namely coherent superpositions of position eigenstates); or, more generally, as localization in Hilbert space of coherent superpositions of many-body Fock states, leading to the recently discovered phenomenon of many-body localization \cite{NandkishoreH2015,AletL2018,Abaninetal2019}. When focusing one's attention on the low-energy physics of many-body systems, localization is generally characterized by the disruption of long-range correlations in the ground state via a quantum phase transition -- such as the superfluid-insulator transition \cite{Fisheretal1989}, the Aubry-Andr\'e transition \cite{AubryA1980}, the superconductor-insulator transition \cite{Trivedi-book}, and the Mott-glass transition \cite{VojtaT2019}. The transition is moreover accompanied by the localization of elementary excitations above the ground states \cite{Alvarez-ZunigaL2013, Vojta2013}, as revealed \emph{e.g.} by the dynamical structure factor. In the case of systems breaking a continuous symmetry (such as bosonic or fermionic superfluids; spin models with U(1) or SU(2) symmetry, etc.) the extended nature of the excitations is protected by the broken symmetry via the Goldstone theorem \cite{Auerbachbook}, so that a quantum phase transition restoring the symmetry is a necessary condition for the localization of elementary excitations down to the lowest energies.  When both the ground state and the excitation spectrum undergo localization, one expects a drastic transition both in the low-energy dynamics (fundamentally governed by the nature of the excitations) as well as in the low-temperature thermodynamics (fundamentally controlled by the ground-state properties). 
  
  \begin{center}
\begin{figure}[ht!]
\includegraphics[width=\columnwidth]{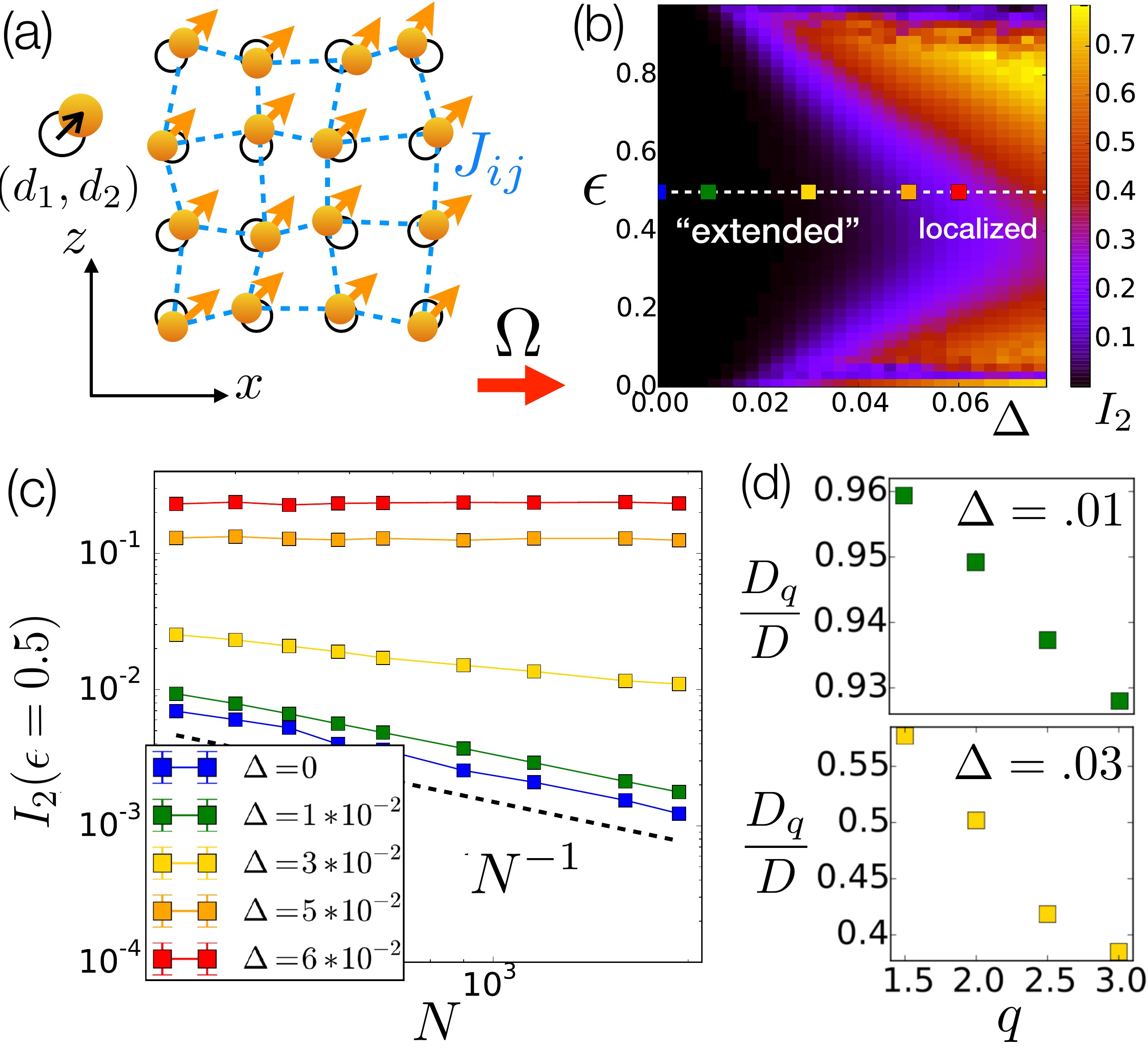}
\caption{\emph{Positionally disordered 2$D$ spin array}. (a) Sketch of the system geometry; the blue dashed lines indicate the couplings (limited to nearest neighbors for illustration purposes), and the spin orientations offer a sketch of the mean-field ground state; (b) Inverse participation ratio $I_2$ of the harmonic eigenmodes as a function of disorder and normalized eigenmode energy $\epsilon$ (see text) for a system size of $L=40$; (c) scaling of $I_2$ of the eigenmodes with $\epsilon=0.5$ for various disorder strengths (referred to by points with same colors in panel (b)); (d) effective fractal dimension $D_q$ of the $\epsilon=0.5$ eigenmodes.}
\label{f.fig1}
\end{figure}
\end{center}
  
   On the other hand, in the case of phases breaking a discrete symmetry (of interest to this work) disorder may have a minimal effect on the ground-state correlations, leaving the symmetry broken, and yet lead to localization (partial or total) of the elementary excitations. This fact is seldom recognized in the literature, although a few examples have been reported in the case of (quasi-)disordered quantum Ising models \cite{Igloietal2012,ChandranL2017,Crowleyetal2018}. In this Letter we present a realistic model, capturing the physics of two-dimensional Rydberg-atom arrays \cite{Barredoetal2017, Lienhardetal2018, Guardado-Sanchezetal2018}, which precisely manifests this phenomenon. Our model exhibits an ordered ground state with very weak quantum fluctuations, coexisting with a rich structure in the linearized low-energy spectrum: the latter comprises a regime of seemingly extended excitations, a regime with localized and extended excitations (exhibiting a mobility edge), and a regime of fully localized excitations. Our main finding is that the fundamental difference between the nature of the ground state and that of the elementary excitations causes a radical dichotomy between the dynamics and the thermodynamics of the system. Indeed the disorder has minor effects on the thermodynamics (in which long-range order is protected by a finite energy gap, and persists up to a finite critical temperature). But the closed-system dynamics, following a low-energy quantum quench, is instead radically different from that of the system without disorder: instead of the usual light-cone propagation of correlations and entanglement found in non-disordered systems \cite{CalabreseC2016}, the model exhibits a unitary dynamics continuously interpolating between ballistic (for an extended spectrum) and arrested behavior (for a fully localized spectrum), with a disorder-dependent critical exponent $z$ governing the power-law growth of entanglement. We conjecture that the anomalously diffusive nature of the dynamics of the system is related to the multifractal nature of the extended excitations in the system at intermediate disorder - which are therefore of the non-ergodic type \cite{fractal-book,EversM2008,Birolietal2012,DeLucaetal2014,Kravtsovetal2015}, having effective support on a subset of lattice sites of zero measure.
 
 \emph{Model Hamiltonian.} The Hamiltonian describing the internal state dynamics in an array of atoms, encoding $S=1/2$ spins in the ground and Rydberg state, reads \cite{Browaeysetal2016}
\begin{equation}
{\cal H} = \sum_{i<j} J_{ij} S^z_i S^z_j  - \sum_i \left [ ~\Omega ~S_i^x + (\delta_0 + \delta_i) ~S^z_i ~\right ]~. 
\label{e.Ham}
\end{equation}
Here the indices $i$ and $j$ run over the sites of a $L\times L$ deformed square lattice (see below) with periodic boundary conditions; $J_{ij} = J_0 |\bm r_i - \bm r_j|^{-6}$ is the van der Waals (vdW) spin-spin interaction between two sites at positions ${\bm r_i}$ and ${\bm r_j}$, $\Omega$ is the Rabi frequency of the field driving the Rydberg transition and detuned by $\delta_0$ away from resonance, while $\delta_i = \sum_j J_{ij}$ is a local shift of the detuning, stemming from the fact that the vdW interaction is not a particle-hole symmetric function of the density of Rydberg excitations \cite{Browaeysetal2016}. In order to simplify the theoretical model, in the following we shall assume that $\delta_0 + \delta_i \approx 0$ \footnote{This choice in fact reduces the amount of disorder in the system, leaving it only on the spin-spin couplings. We have verified that including the detuning disorder in the system does not alter substantially the picture.}. Moreover throughout this work we choose  $\Omega = |J_0|$. The above model Eq.~\eqref{e.Ham} realizes then a quantum Ising model, in which disorder can be simply introduced \emph{geometrically}, namely by randomly positioning the atoms, and automatically creating a (correlated) randomization of the couplings $J_{ij}$. In order to tune the strength of disorder, we choose the spin positions to be of the form ${\bm r_i} = (n_1 + d_1) {\bm e}_1 + (n_2 + d_2) {\bm e}_2$, where ${\bm e}_1$ and ${\bm e}_2$ are orthogonal unit vectors, $n_{1(2)}$ are integers, and $d_{1(2)}$ are random variables uniformly distributed in the interval $[-\Delta,\Delta]$. In this way the spins form a square lattice of unit lattice spacing for $\Delta = 0$, while the lattice is gradually disordered as $\Delta$ grows up (see Fig.~\ref{f.fig1}(a) for a sketch). Hence our model interpolates continuously between a perfect array and a completely random array. The $r^{-6}$ dependence of the coupling on the spin-spin distances translates positional disorder into coupling disorder \cite{SuppMat}. The long-range nature of the couplings creates an additional source of randomness with respect to a simple model of random bonds connecting nearest neighbors; yet the long-range nature of the couplings is not found to be essential to our observations -- we see the same phenomenology upon truncating the interactions to nearest neighbors \cite{SuppMat}. Another remark is in order concerning the sign of the couplings. The Hamiltonian naturally implemented by Rydberg atoms is \emph{antiferromagnetic} ($J_0>0$); yet the central focus of our study is non-equilibrium dynamics, for which the sign of the coupling is irrelevant as long as the initial state is time-reversal invariant (namely, if it has real coefficients on the same basis on which the Hamiltonian is represented as a real matrix) \cite{Frerotetal2018, MenuR2018}. Most importantly, natural initial states for the experimental dynamics are factorized states that are homogeneous, namely they are close to the ground state of the \emph{ferromagnetic} model ($J_0<0$): choosing to evolve them with the ferromagnetic Hamiltonian has the relevant bonus of allowing us to use linearized spin-wave (LSW) theory, since only the low-energy part of the spectrum is explored. For this reason we shall specialize to the case $J_0 < 0$ in the following.    
 
\emph{Spin-wave spectrum: from extended to localized modes.} Our real-space LSW approach is based on three steps: 1) the ground state of the Hamiltonian Eq.~\eqref{e.Ham} is calculated within the mean-field (MF) approximation, namely in the factorized form $|\Psi_{\rm MF}\rangle = \otimes_i \left [ \cos(\theta_i/2) |\uparrow_i\rangle + \sin(\theta_i/2) |\downarrow_i \rangle \right ]$; 2) rotating the local spin operators by an angle $\theta_i$ around the $y$ axis, and mapping the rotated spins onto Holstein-Primakoff (HP) bosons $b_i, b_i^\dagger$ \cite{Auerbachbook}, leads to an Hamiltonian of the form ${\cal H} \approx E_{\rm MF} + \sum_{ij} (b_i^{\dagger}, b_i) {\cal A}_{ij} (b_j, b_j^{\dagger})^{T}$ where $E_{\rm MF}$ is the mean-field ground-state energy, and all terms beyond quadratic have been dropped under the assumption of a diluted gas of HP bosons, $r = (1/N)\sum_i \langle n_i \rangle /(2S) \ll 1$; 3) the linearized Hamiltonian is then diagonalized via the Bogoliubov transformation $b_i = \sum_\alpha ( u_i^{(\alpha)} \beta_\alpha+ v_i^{(\alpha)} \beta_\alpha^\dagger)$, where $\beta_\alpha, \beta_\alpha^\dagger$ are bosonic operators such that the linearized Hamiltonian takes the diagonal form, ${\cal H} \approx E_{\rm MF} + \sum_\alpha \varepsilon_\alpha \beta_\alpha^\dagger \beta_\alpha$. The vector $(\{u^{(\alpha)}_i\},\{ v^{(\alpha)}_i \})$, normalized as $\sum_i [ (u_i^{(\alpha)})^2 -(v_i^{(\alpha)})^2 ] = 1$, provides the spatial profile of the $\alpha$-th eigenmode with eigenenergy $\varepsilon_{\alpha}$.  Our calculations apply to $L\times L$ lattices with periodic boundary conditions and $L$ up to 50. All data shown were averaged over several (typically 25) realizations of disorder; increasing the disordered statistics (as we have done in selected cases up to 100 realizations) does not alter the results. 

\begin{center}
\begin{figure*}[ht!]
\includegraphics[width=\textwidth]{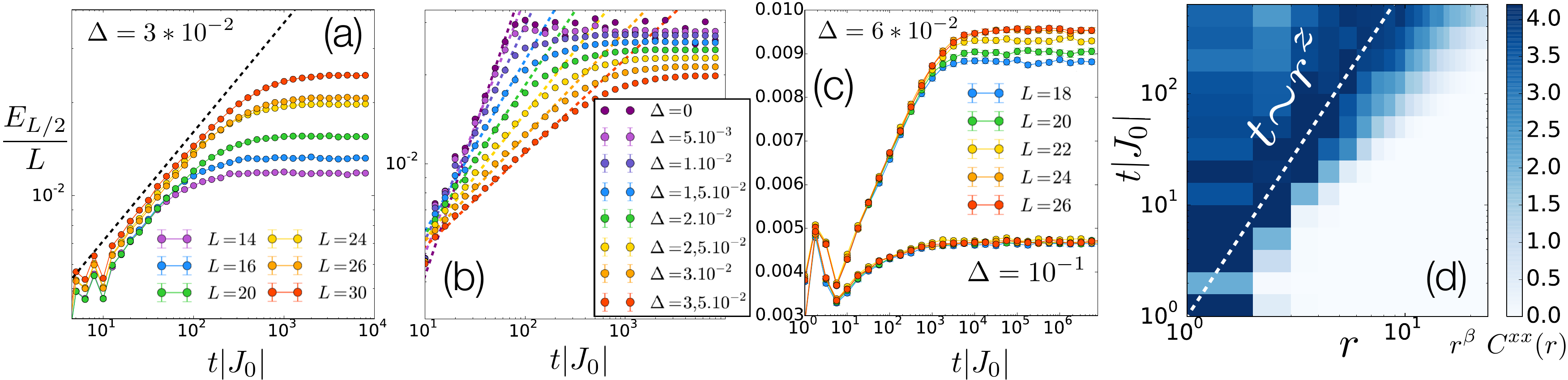}
\caption{\emph{Dynamics in the positionally disordered spin array.} Entanglement entropy (EE) of the half torus $L/2\times L$ for $\Delta = 3*10^{-2}$ and various system sizes, showing consistently a power-law scaling $E_{L/2} \sim t^{1/z}$ with $z\approx 2.8$; (b) EE growth for different disorder strengths in the anomalously diffusive regime - here $L=30$; (c) EE growth in the localized regime;  (d) spreading of correlations for the $x$ spin components, multiplied $r^\beta$ where $\beta \approx 1.1$ is the exponent of the asymptotic power-law decay -- here $\Delta = 3 * 10^{-2}$ and $z$ as in (a).}
\label{f.fig2}
\end{figure*}
\end{center}

The localization properties of the LSW eigenmodes can be monitored by studying the generalized inverse participation ratio (IPR) of the mode profiles, $I_q^{(\alpha)} = \sum_i [ (u_i^{(\alpha)})^2 -(v_i^{(\alpha)})^2 ]^q$. \footnote{This expression extends the generalized inverse participation ratio of normalized lattice eigenfunctions $\psi_i$, $I_q = \sum_i |\psi_i|^{2q}$ \cite{fractal-book,EversM2008}, to Bogolyubov eigenmodes. Elevating the difference $D_i^{(\alpha)} = (u_i^{(\alpha)})^2 -(v_i^{(\alpha)})^2$ to a potentially non-integer power $q$ does not raise any problem, because for all of the quadratic Hamiltonians we considered $D_i^{\alpha}\geq 0$. This is a consequence of the diluteness of the gas of Holstein-Primakoff bosons in the ground state - underlying the correctness of the quadratic approximation - since 
${\langle n_i \rangle} = \sum_\alpha (v_i^{(\alpha)})^2 \ll 1$ implies \emph{a fortiori} $(v_i^{(\alpha)})^2 \sim {\langle n_i \rangle}/N  \ll 1$ for any $\alpha$, while $(u_i^{(\alpha)})^2 \sim 1/N \gg  (v_i^{(\alpha)})^2$ for the mode to be correctly normalized, $\sum_i D_i^{(\alpha)} = 1$.}
 For $q=2$ one recovers the conventional IPR ($I_2$), providing the inverse volume of the effective spatial support of the mode. An $I_2$ scaling to a finite value in the thermodynamic limit indicates a localized mode, while an $I_2$ scaling to zero as $L^{-D}$ (with $D$ the number of dimensions) signals an extended mode; on the other hand an $I_2$ showing an intermediate scaling as $L^{-D_2}$ with $D_2<D$ signals the extended non-ergodic nature of a mode \cite{fractal-book,EversM2008}.  Fig.~\ref{f.fig1}(b) shows $I_2$ as a function of disorder strength $\Delta$ and energy density $\epsilon_\alpha = (\varepsilon_\alpha - \varepsilon_{\rm min})/(\varepsilon_{\rm max} - \varepsilon_{\rm min})$ (where $\varepsilon_{\rm min(max)}$ are the two extrema of the LSW spectrum) for the model Eq.~\eqref{e.Ham} with $\Omega = |J_0|$; while Fig.~\ref{f.fig1}(c) shows the scaling of $I_2$ with system size for modes in the center of the spectrum ($\epsilon = 0.5$). In both graphs $I_2$ is averaged over groups of modes that lie in the same $\epsilon$ interval (of width $0.02$), and further averaged over disorder realizations. There we observe three regimes: I) for weak disorder ($\Delta \lesssim 10^{-2}$) all modes are extended; II) for intermediate disorder ($10^{-2} \lesssim \Delta \lesssim 4*10^{-2}$) low- and high-energy modes display a scaling of $I_2$ consistent with localized behavior \cite{SuppMat}, but modes in the central region of the spectrum suggest that $I_2$ scales to zero (as in the case of extended modes), possibly with a multifractal scaling spectrum -- see below; two mobility edges (between localized and non-ergodic extended states) may therefore appear in the spectrum; III) for strong disorder ($\Delta \gtrsim 4*10^{-2}$) all modes undergo localization. 
 
 A closer look at the nature of the in-between regime II can be obtained by looking at the size and $q$-index dependence of the $I_q$'s, which are in general expected to obey the scaling $I_q \sim L^{-D_q(q-1)}$, where $D_q$ provides the effective dimension of the modes' support. For localized states $D_q = 0$, while $D_q = D$ for extended states, independently of $q$; a $q$ dependence of the effective dimension signals instead multifractality \cite{fractal-book,EversM2008}. Fig.~\ref{f.fig1}(d) shows that in regime II the exponent $D_q$ for the mid-spectrum modes ($\epsilon = 0.5$), extracted from power-law fits on the size scaling of $I_q$ with $L=18\div34$ (see \cite{SuppMat}), takes indeed a $q$ dependent value, suggesting the multifractal and non-ergodic nature of the modes themselves.   
  
A few comments are in order at this point. The existence of a localization transition at finite disorder in systems with $D\leq 2$ and short-range hoppings is excluded in general for quadratic Hamiltonians in the orthogonal symmetry class, although it is known to exist in other (fermionic) Hamiltonians belonging to different symmetry classes \cite{fractal-book,EversM2008,Fisher1992, YoungR1996, Igloietal2012}. In the case of bosonic quadratic Hamiltonians including hopping and pairing terms -- such as the one of interest here -- a classification of possible Anderson transitions is still absent, leaving the possibility of the existence of such a transition in $D=2$, as also suggested by recent numerical studies \cite{AlvarezZunigaL2013}. An alternative scenario to the one we propose here would be that of eigenmodes which are localized as soon as disorder is switched on, yet with an exceedingly large localization length $\xi$: under this assumption, the physics of states that are characterized here as extended applies in fact to length-scales strictly smaller than $\xi$, and to time scales smaller than the time necessary for quasi-particles to cover distances comparable with $\xi$ \cite{Sebbahetal1993}.
In our specific case, a most striking aspect is that the putative localization transition in the Bogolyubov eigenmode spectrum occurs for a very weak disorder, due to the strong dependence of the vdW couplings on distance \cite{SuppMat}; and it is \emph{not} accompanied by any concomitant ground-state phase transition. In fact the quantum renormalization of the mean-field order parameter is extremely weak throughout the localization transition, namely $r \ll 1$ \cite{SuppMat}, and therefore the ground state retains its long-range ordered nature, protected by a large energy gap. This result fully vindicates the LSW approach in the parameter range explored here, and it invites us to use the same approach to investigate the non-equilibrium dynamics at low energy.

\emph{Quench dynamics: from anomalous diffusion to localization.} A natural way to probe the nature (extended vs. localized) of the Bogoliubov eigenmodes, for theory and experiments alike, is to study the non-equilibrium dynamics following the initialization of the system in a non-stationary state (quantum quench). Here we specialize our attention to quenches that initialize the system in the MF ground state  $|\Psi_{\rm MF}\rangle$, corresponding to the vacuum of HP bosons: the subsequent dynamics generates entanglement and correlations among the sites by the creation/destruction and propagation of pairs of HP bosons. Nonetheless the gas of HP bosons generated dynamically remains very diluted ($r \ll 1$), as the magnetic order in the ground state is protected up to a finite critical temperature / energy density; and preparing the system in the MF ground state leaves the system well below that energy density. 
 
  We monitor the post-quench evolution of the system based on the evolution of entanglement and correlations. Figs.~\ref{f.fig2}(a-c) show  the spreading of entanglement, quantified as the von Neumann entanglement entropy (EE) of half of the system $E_{L/2}$ (a half torus of size $L/2\times L$) calculated as in Refs.~\cite{FrerotR2015,Frerotetal2017, Frerotetal2018}. In general the EE at any finite time obeys and area-law scaling $S \sim L$, due to the fact that ${\cal O}(L)$ quasiparticles per unit time propagate across the interface between the two subsystems. The EE further grows in time due to the propagation of quasiparticles within each subsystem: in the case of unarrested quasiparticle propagation, the EE grows up to a value satisfying a volume law scaling, $E_{L/2} \sim L^2$, and it does so in a characteristic time $t_L^{*}$ generally scaling as a power-law of the subsystem size $L^z$ -- namely $E_{L/2} $ grows as $t^{1/z}$ before reaching saturation.  In the absence of disorder ($\Delta = 0$), the ballistic transport of correlated pairs of quasiparticles leads to linear growth of EE, $z=1$,  and $t_L^* \approx L/(2v)$ where $v$ is the maximum group velocity of the Bogolyubov excitations \cite{Cevolanietal2016,Frerotetal2018}. On the other hand, as soon as disorder is introduced in the system, the data exhibit a dynamical exponent $z$ which is continuously growing with disorder, spanning the whole range from sub-ballistic ($1<z<2$) to sub-diffusive ($z>2$) dynamics as $\Delta$ increases (Fig.~\ref{f.fig2}(b) and \cite{SuppMat}). The exponent $z$ appears to diverge at the localization transition of the quasi-particle spectrum ($\Delta \approx 5*10^{-2}$); immediately beyond that value (for $\Delta = 6*10^{-2}$), logarithmic growth of EE seems to set in for the system sizes that we could access; but for stronger disorder ($\Delta = 10^{-1}$) the EE spreading is clearly arrested at long times (see Fig.~\ref{f.fig2}(c)), and the asymptotic (pre-thermalized) state retains an area-law scaling.    
  
  The observed growth of entanglement entropy is fully consistent with the spreading of correlations, whose front advances as well in a power law fashion as $r\sim t^{1/z}$ with an exponent $z$ consistent with the one governing the growth of EE -- see Fig.~\ref{f.fig2}(d). This observation makes the anomalous $z$ exponent governing the dynamics immediately accessible to experiments equipped with single-site resolution. Below the localization transition of the spectrum the correlation function relaxes towards an asymptotic profile which displays a \emph{power-law} decay as $r^{-\beta}$ with $\beta \ll 6$ -- at odds with a thermal state, in which correlations should decay as $r^{-6}$, stabilized by the long-range interactions \cite{Frerotetal2017}. This aspect singles out the non-equilibrium character of the \emph{pre}-thermal state emerging from the dephasing dynamics of non-interacting spin waves. Finally in the localized regime the correlation-front dynamics is much slower (possibly logarithmic), converging to an exponentially decaying asymptotic correlation in the pre-thermalized state \cite{SuppMat}. Very similar dynamical behavior (anomalous diffusion at intermediate disorder, localization at strong disorder) is also found for the spreading of an initially localized spin flip on top of the MF ground state, namely for the evolution of the state $b_i^{\dagger} |\Psi_{\rm MF}\rangle$. 
  
 \emph{Conclusions.} We unveil a very complex quantum dynamics of elementary excitations in positionally disordered quantum Ising models with vdW interactions. For the sizes (up to $50^2$ sites) we explored, the system exhibits anomalous diffusion with continuously varying dynamical exponent for a broad range of disorder strengths, related to the multifractal nature of the quasiparticle eigenmodes; as well as arrested dynamics and localization for strong disorder. In disordered tight-binding models, anomalous diffusion and multifractality typically require fine tuning of the system at the Anderson transition \cite{fractal-book,EversM2008,OhtsukiK1997}; multifractality of single-particle eigenstates over extended energy/parameter ranges has been reported in tight-binding models on disordered hierarchical lattices \cite{TikhonovM2016,Sonneretal2017,BiroliT2018}, on positionally disordered 3D systems with dipolar hoppings \cite{Dengetal2016},  as well as in random-matrix models \cite{Kravtsovetal2015}; multifractality and anomalous diffusion are also exhibited by tight-binding models on quasi-crystals \cite{Passaroetal1992,ZhongM1995}. Here we show that multifractality and anomalous diffusion appear to be generic properties of the linearized many-body excitations and dynamics (respectively) in a positionally disordered Ising model, respectively, highlighting therefore its theoretical interest. Most importantly, this model is readily implemented experimentally with Rydberg atoms, and by design it interpolates between the case of atoms trapped in regular arrays of individual traps \cite{Labuhnetal2016, Endresetal2016, Barredoetal2017, Bernienetal2017, Lienhardetal2018, Guardado-Sanchezetal2018} and atoms trapped by global confining potentials, whose position can be considered as frozen on the time scale of internal-state dynamics \cite{PineiroOriolietal2018, Whitlocketal2018}. Even in the case of atoms trapped in regular arrays, recent experiments have already pointed out the existence of positional disorder due to thermal fluctuations of the atom positions within their individual traps  \cite{Marcuzzietal2017} and its possible role in the dynamics. Here we show that even a very weak positional disorder (with relative fluctuations of a few percents, comparable to those already reported in experiments \cite{Marcuzzietal2017}) can lead to subdiffusion and localization in the low-energy dynamics of the system; this aspect may play a role in the quantum simulation of quantum magnetism via the (quasi-)adiabatic preparation of Rydberg-atom arrays \cite{Lienhardetal2018}.  

\emph{Acknowledgements.} We thank A. Jagannathan and A. Mirlin for useful discussions. Numerical computations were performed on the PSMN cluster (ENS Lyon). This work is supported by ANR (``EELS" project).

\vspace{1cm} 
 
\begin{center}
{\bf Supplementary Material} \\
{\bf \emph{Anomalous diffusion and localization in a positionally disordered quantum spin array}}
\end{center}

\setcounter{figure}{0}

\section{Ground-state order parameter for increasing disorder}   

\begin{center}
\begin{figure}[ht!]
\includegraphics[width=0.8\columnwidth]{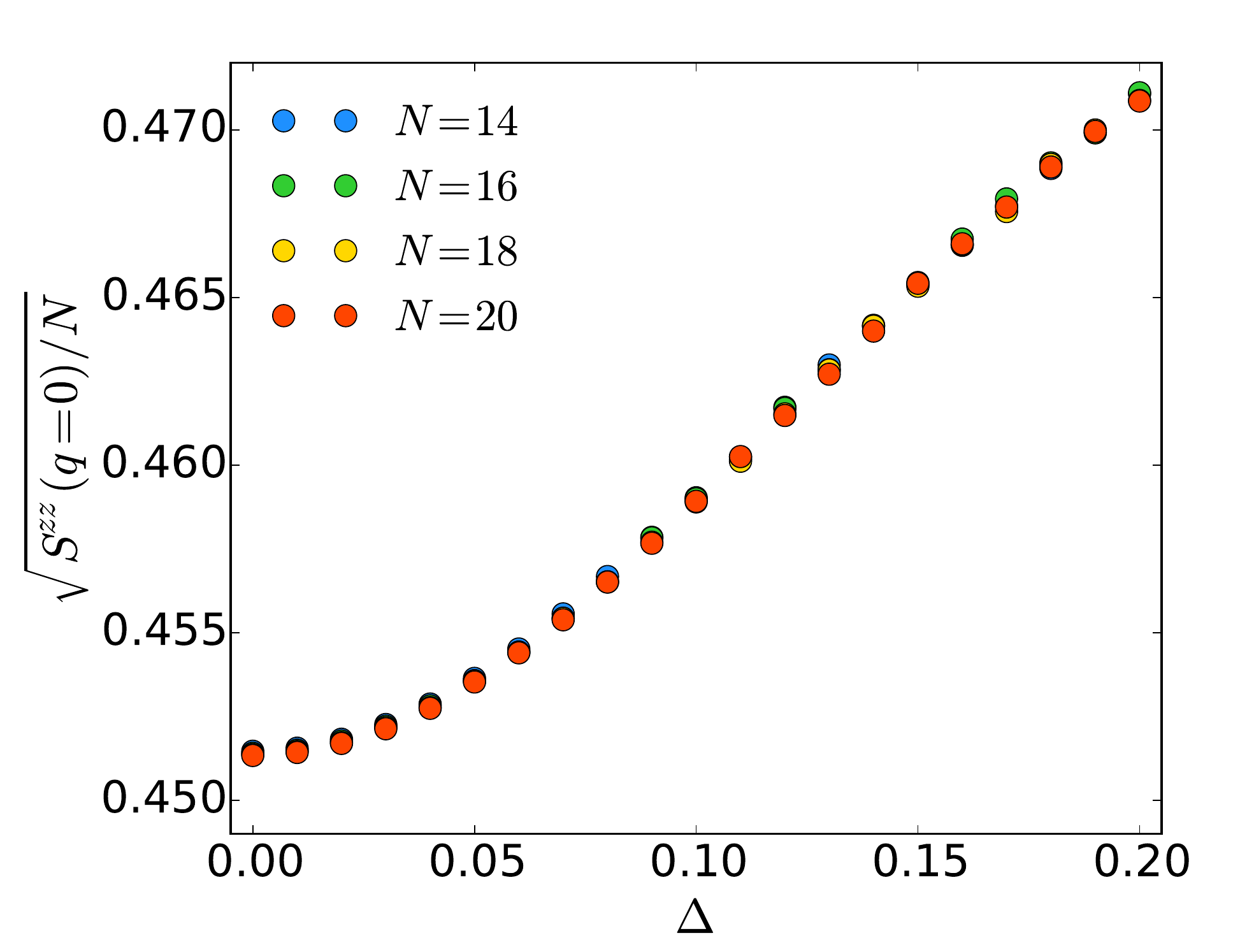}
\caption{Ising order parameter in the spin-wave ground state, plotted as a function of disorder and for different system sizes.}
\label{f.structfact}
\end{figure}
\end{center}

Fig.~\ref{f.structfact} shows the Ising order parameter of the spin-wave ground state, obtained through the structure factor
\begin{equation}
m_0 = \sqrt{ \frac{S^{zz}(q=0)}{N}}
\end{equation}
where $S^{zz}(q=0) = (1/N)~\sum_{ij} \langle S_i^z S_j^z \rangle $. We observe that the ferromagnetic nature of the ground state is 
\emph{strengthened} by disorder, as a consequence of the increase of the average coupling strength. Indeed, because of the convex nature of the $r^{-6}$ function, it is immediate to see that $[ r^{-6} ]_{\rm av} > [r]^{-6}_{\rm av}$ where $[...]_{\rm av}$ denotes here the average over disorder.

\section{Probability distribution of spin-spin couplings}

 Here we discuss analytically the simple case of the coupling between two atoms in one spatial dimension, whose positions are displaced by two independent random displacements $\delta_1$ and $\delta_2$ uniformly distributed in the interval $[-\Delta, \Delta]$. We shall consider a generic power-law coupling, whose dimensionless form is 
 \begin{equation}
 y = \frac{J}{J_0} = \frac{1}{|1-\delta_1 + \delta_2|^\alpha}~.
 \end{equation}
 The difference of the displacements, $z = \delta_1 - \delta_2$, is a random variable distributed according to a triangle distribution
 \begin{equation}
 p(z) = \theta(-z) (z+2\Delta) + \theta(z) (2\Delta-z)
 \end{equation}
 with support on the interval $[-2\Delta, 2\Delta]$ -- here $\theta(x)$ is the Heaviside function. The resulting distribution for the $y = y(z)$ variable, defined between $y_{+} = |1+2\Delta|^{-\alpha}$ and $y_{-} = |1-2\Delta|^{-\alpha}$, is then
 \begin{eqnarray}
 P_\Delta(y) =  \frac{1}{4\alpha \Delta^2 y^{1+1/\alpha}} & \Big [ &\theta (y-1) \left (y^{-1/\alpha} + 2\Delta -1\right) \nonumber \\
  &+&  \theta(1-y) \left (2\Delta + 1 - y^{-1/\alpha} \right ) \Big ]~~~~
 \end{eqnarray}
whose $n$-th moment takes the form
\begin{equation}
[ y^n ]_{\rm av} = \frac{(1-2\Delta)^{2-\alpha n}+(1+2\Delta)^{2-\alpha n}-2}{4\alpha^2 \Delta^2 (n-2/\alpha)(n-1/\alpha)}
\end{equation}
 which, for small displacements, $\Delta \to 0$, admits the expansion
 \begin{equation}
 [ y^n ]_{\rm av} \approx 1 + \frac{\alpha^2}{6} n\left (n+\frac{1}{\alpha} \right ) \Delta^2 + {\cal O} (\Delta^4)~.
\end{equation}
The second moment has then the following behavior for small displacements: 
\begin{equation}
{\rm Var}(y) \approx \frac{\alpha^2}{3}  \Delta^2 + {\cal O} (\Delta^4) =  \frac{\alpha^2}{2} ~{\rm Var}(\delta) + {\cal O} ({\rm Var}(\delta)^2) ~.
\end{equation}
Hence we see that for $\alpha > \sqrt{3}$ the variance of the spin-spin coupling is amplified with respect to the variance of the particle displacements by a factor of $\alpha^2/3$. This becomes particularly severe in the case of var-der-Waals interactions with $\alpha=6$. 

\begin{center}
\begin{figure}[ht!]
\includegraphics[width=0.9\columnwidth]{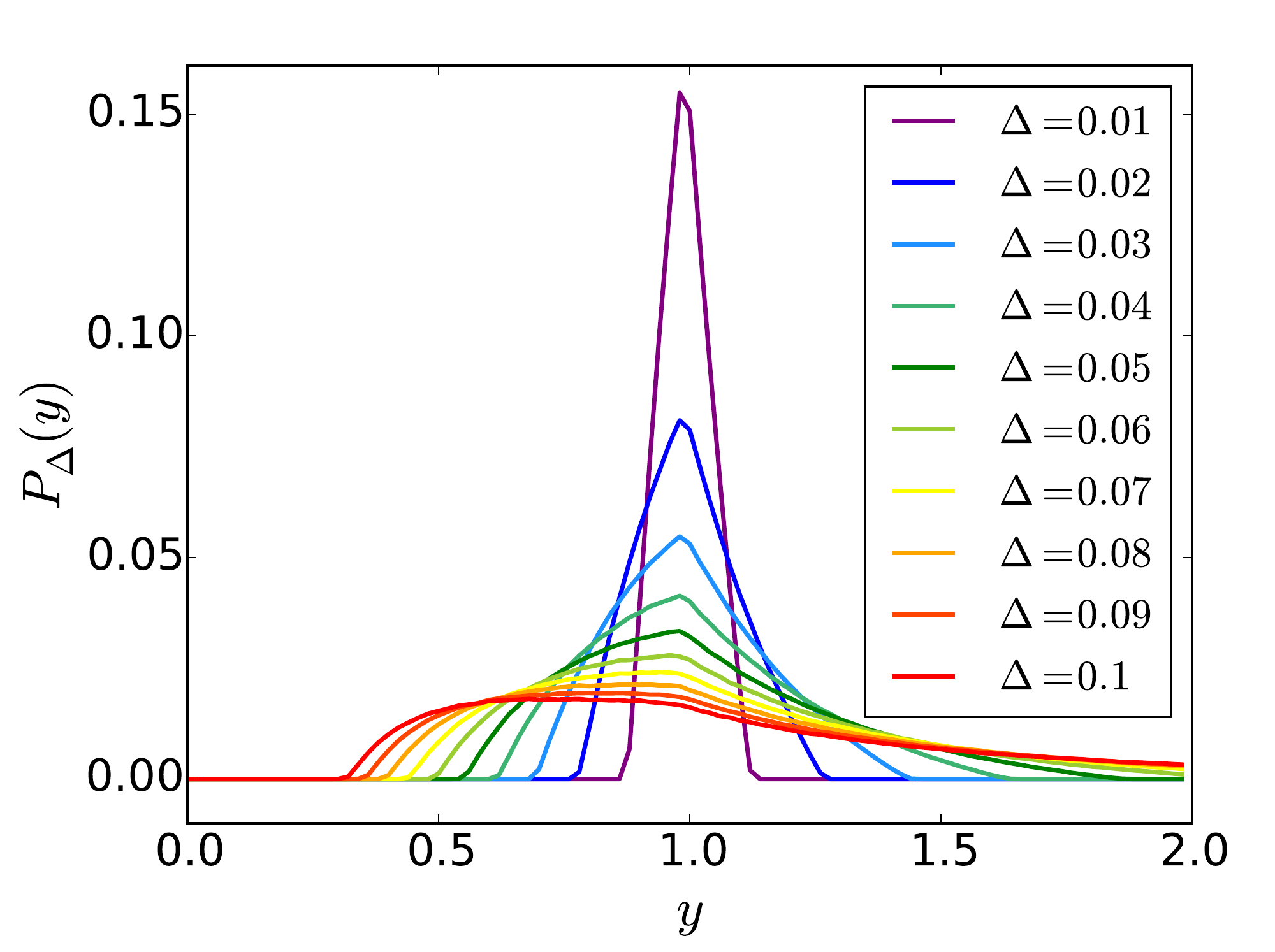}
\caption{Probability distribution of the nearest-neighbor couplings on the square lattice for various values of disorder.}
\label{f.PDFcouplings}
\end{figure}
\end{center}

\begin{center}
\begin{figure}[ht!]
\includegraphics[width=\columnwidth]{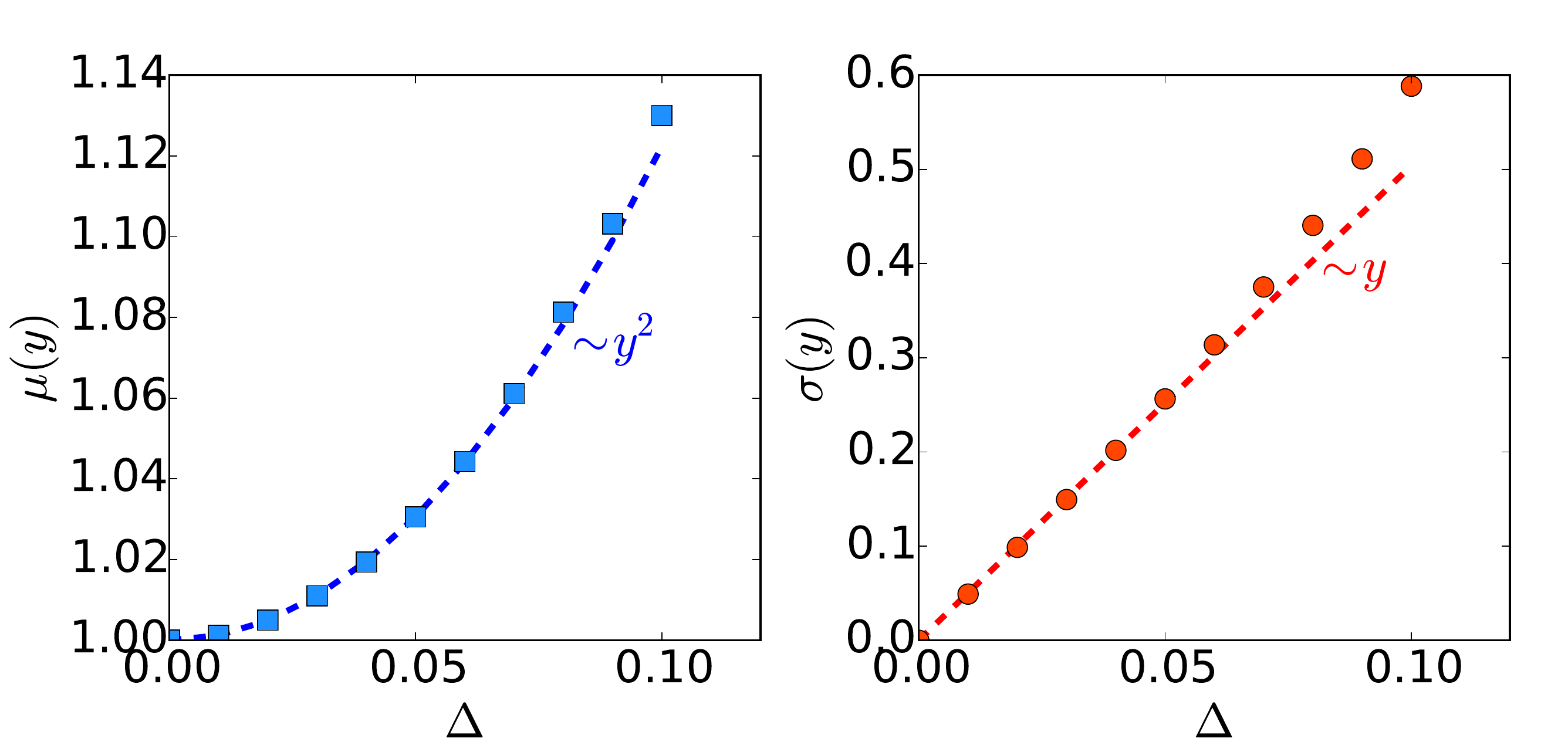}
\caption{First two moments of the coupling distribution as a function of disorder: average (left panel) and standard deviation (right panel).}
\label{f.PDFmomenta}
\end{figure}
\end{center}

The actual dependence of the probability distribution $P(y)$ for the $D=2$ system studied in this work, and in particular of its first and second moments on the disorder $\Delta$, is shown in Figs.~\ref{f.PDFcouplings} and \ref{f.PDFmomenta}. There we see that the distribution $P_{\Delta}(y)$ has an asymmetric form, with a skewness (not shown) growing linearly with the amount of disorder. The average value grows quadratically with disorder, while the standard deviation grows linearly with $\Delta$, as predicted for $D=1$, but with an even larger proportionality factor $\sigma(y) \approx 5 ~\Delta$ (instead of $\alpha/\sqrt{3} ~\Delta = \sqrt{12} ~\Delta$). This amplification factor provides for instance a $\sigma(y) = \sqrt{{\rm Var}(y)} \sim 0.2$, namely a $20 \%$ fluctuations of the couplings, when $\Delta \sim 0.04$, justifying that full localization of the eigenmodes can take place in the spin-wave spectrum already for particle positions displaying relative fluctuations of a few percents.

\section{Quadratic bosonic Hamiltonian}   
 
 After rotating the spin operators to align the local quantization axis with the local orientation of the spin in the mean-field ground state:
  \begin{eqnarray}
 S'^x_i &=&  \cos\theta_i S^x_i - \sin\theta_i S^z_i \nonumber \\
 S'^y_i &=& S_i^y \nonumber \\
 S'^z_i & = &  \sin\theta_i S^x_i + \cos\theta_i S^z_i   
 \end{eqnarray}
 the spin Hamiltonian takes the form
\begin{eqnarray}
{\cal H} & = & \sum_{ij} J_{ij} \cos\theta_i \cos\theta_j S'^z_i S'^z_j \nonumber \\
& + & \sum_{ij} J_{ij} \cos\theta_i \sin\theta_j S'^z_i S'^x_j \nonumber \\
& + & \sum_{ij} J_{ij} \sin\theta_i \cos\theta_j S'^x_i S'^z_j \nonumber \\
& + & \sum_{ij} J_{ij} \sin\theta_i \sin\theta_j S'^x_i S'^x_j \nonumber \\
& - & \sum_i (\Gamma\cos\theta_i - H_i \sin\theta_i) S'^x_i  \nonumber \\
& - & \sum_i (\Gamma\sin\theta_i + H_i \cos\theta_i) S'^z_i ~.
\label{e.Ham'}
\end{eqnarray}
where here we have also included the presence of a local longitudinal field $H_i$. Introducing the linearized Holstein-Primakoff transformation $S^+ \approx \sqrt{2S} ~b$, $S^z = S - b^\dagger b$, the above Hamiltonian is then reduced to the quadratic form that we quoted in the main text, with the matrix 
${\cal A}_{ij}$ given by
\begin{equation}
{\cal A}_{ij} = \begin{pmatrix} D_i \delta_{ij} + C_{ij} & C_{ij} \\ C_{ij} &   D_j \delta_{ij} + C_{ij} \end{pmatrix}
\end{equation}
where
\begin{eqnarray}
D_i & = & - S \sum_l J_{il} \cos\theta_i \cos\theta_l - \frac{1}{2} \left ( \Gamma \sin\theta_i + H_i \cos\theta_i \right ) \nonumber \\
C_{ij} & = & \frac{J_{ij} S}{2} ~\sin\theta_i \sin\theta_j~.
\end{eqnarray}
The random nature of the couplings $J_{ij}$ and of the angle $\theta_i$ characterizing the mean-field ground state induces randomness on all the entries of the matrix. 

\section{Multifractal scaling of the inverse participation ratio}   
    
\begin{center}
\begin{figure}[ht!!]
\includegraphics[width=0.85\columnwidth]{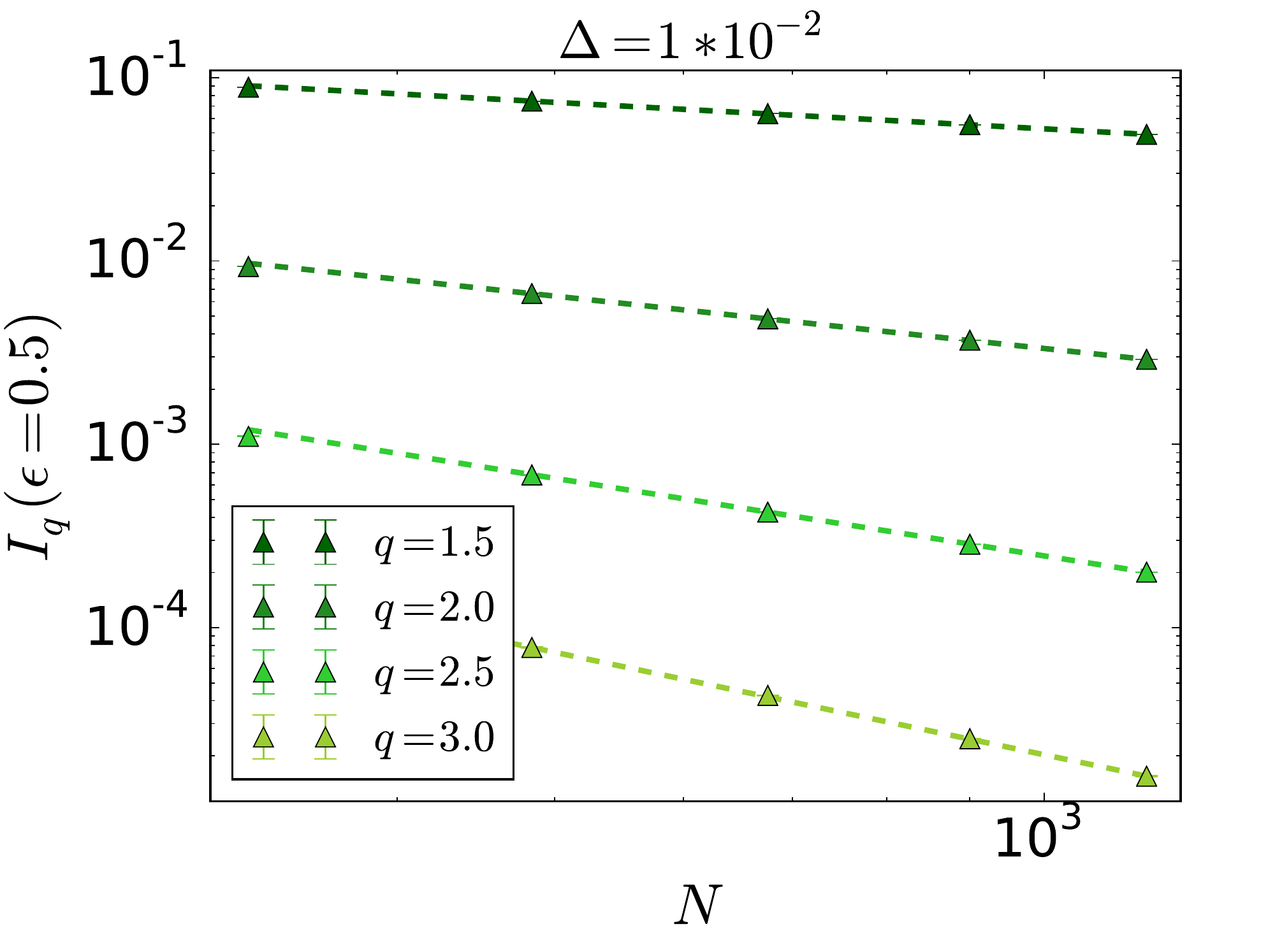}
\includegraphics[width=0.85\columnwidth]{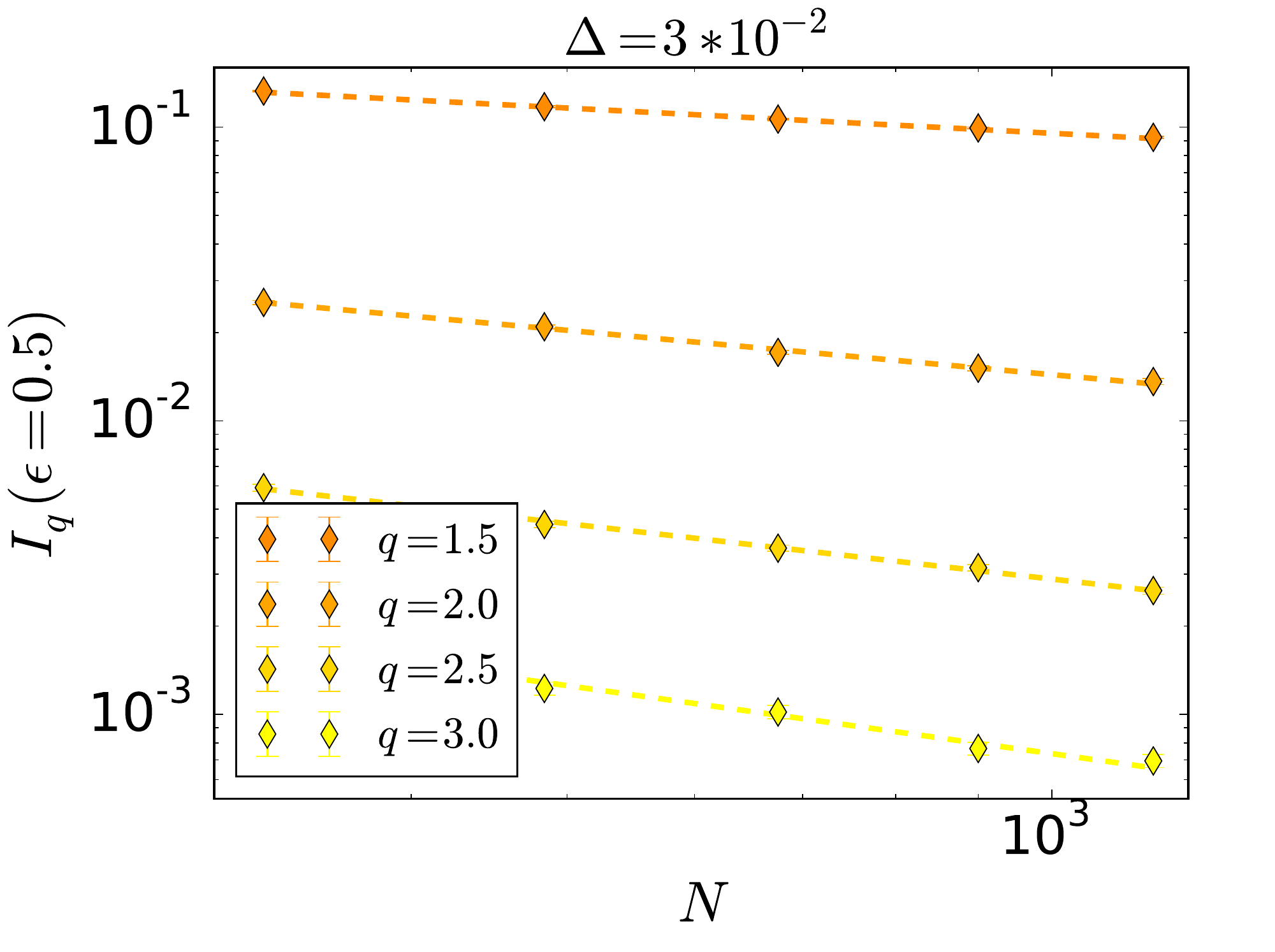}
\includegraphics[width=0.85\columnwidth]{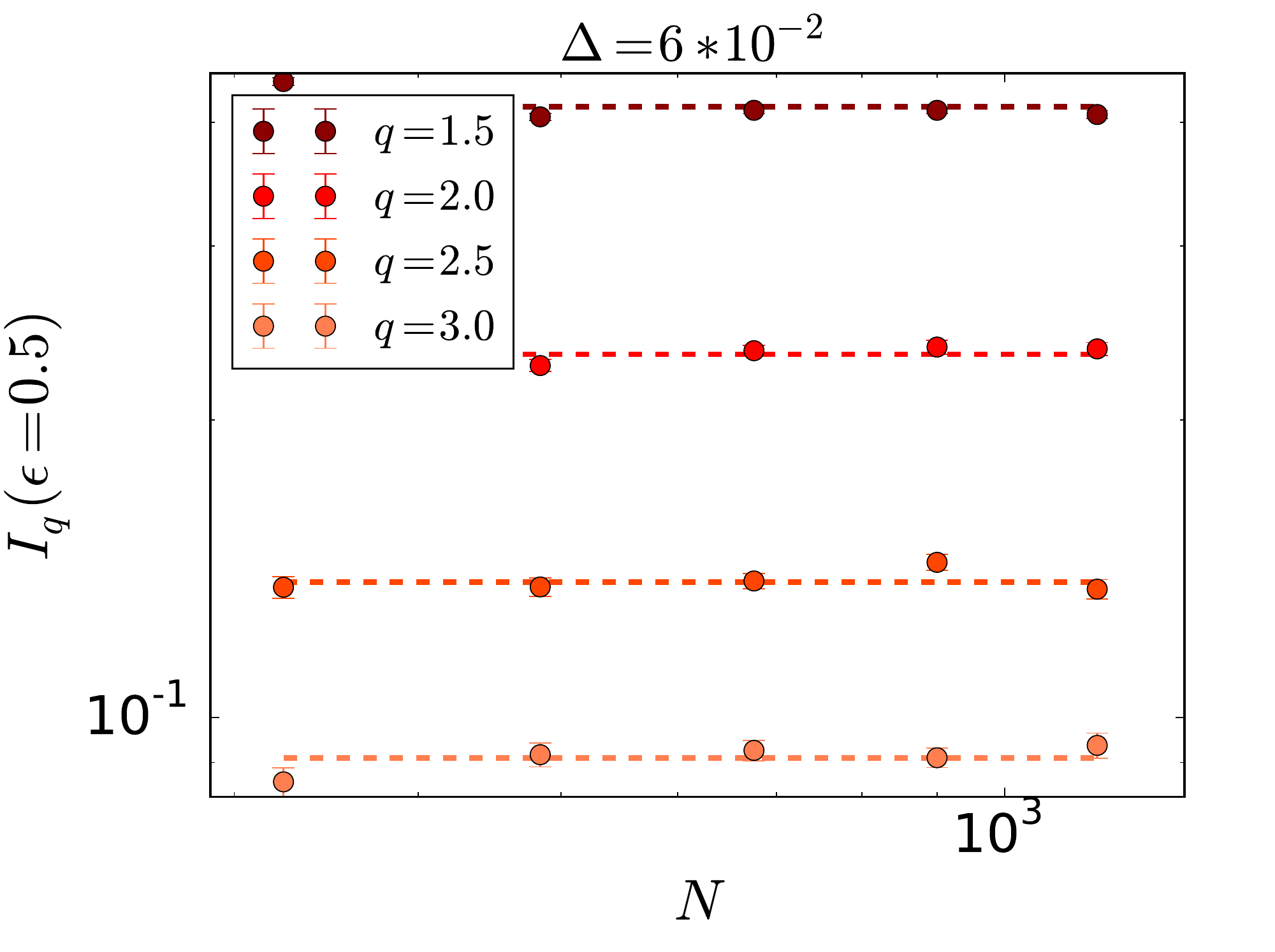}
\caption{Scaling of the generalized inverse participation ratio $I_q$, averaged over eigenmodes in the middle of the spectrum $\epsilon=0.5$, for different disorder strengths $\Delta$. The dashed lines indicate power-law fits in the form $I_q = N^{-D_q(q-1)/D}$; the resulting exponents are shown in Fig.~1 of the main text.} 
\label{f.scaling}
\end{figure}
\end{center}

 Fig.~\ref{f.scaling} shows the size scaling of the generalized inverse participation ratio of the eigenmodes $I_q$ in the center of the spectrum (namely for $\epsilon = 1/2$) and for three disorder values ($\Delta = 10^{-2}$, $3*10^{-2}$ and $6*10^{-2}$). The first two disorder values show a power-law decrease of the inverse participation ratio with a $q$-dependent power-law exponent, suggesting the multifractal nature of the corresponding eigenmodes -- the corresponding $D_q$ exponent is shown in Fig.~1 of the main text. On the other hand for the last value of disorder the scaling is essentially absent, consistent with localized modes. 

\section{Anomalous diffusion for nearest-neighbor interactions}

\begin{center}
\begin{figure}[ht!]
\includegraphics[width=0.8\columnwidth]{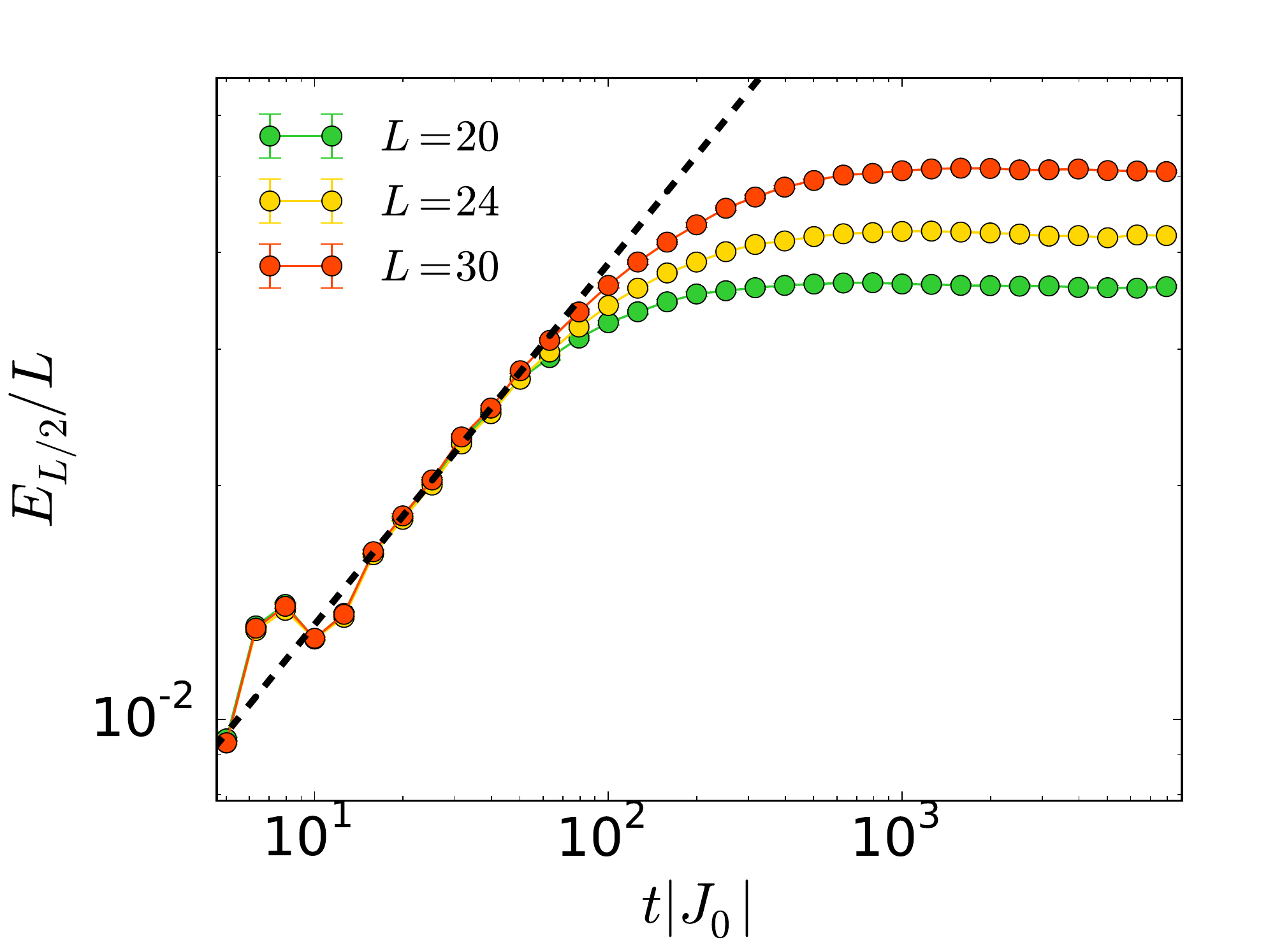}
\caption{Half-system entanglement entropy as a function of time for the model with interactions truncated to nearest neighbor. Here $\Delta = 3*10^{-2}$, and the evolution follows a quench from the mean-field ground state; the dashed line corresponds to a power-law growth as $t^{1/z}$ with  $z\approx 2.15$.}
\label{f.nn}
\end{figure}
\end{center}

An unconventional ingredient of our model of interest, motivated by the experiments on Rydberg atoms, is the presence of van-der-Waals interactions decaying as $r^{-6}$. Here we show that the existence of a power-law tail in the interactions is in fact inessential to the main observation of our work, namely the existence of anomalous diffusion related to multifractality of the linearized eigenmodes. 
Fig.~\ref{f.nn} shows the time dependence of the entanglement entropy (EE) of the half torus following a quench from the mean-field ground state for a model in which the disordered interactions have been truncated to nearest neighbors. There we observe that the EE show a consistent power-law growth with time as $t^{1/z}$, over a time range which increases as the system size is increased, and with an anomalous diffusion exponent of $z\approx 2.15$. 

\section{Dynamical exponent as a function of disorder}

\begin{center}
\begin{figure}[ht!]
\includegraphics[width=0.8\columnwidth]{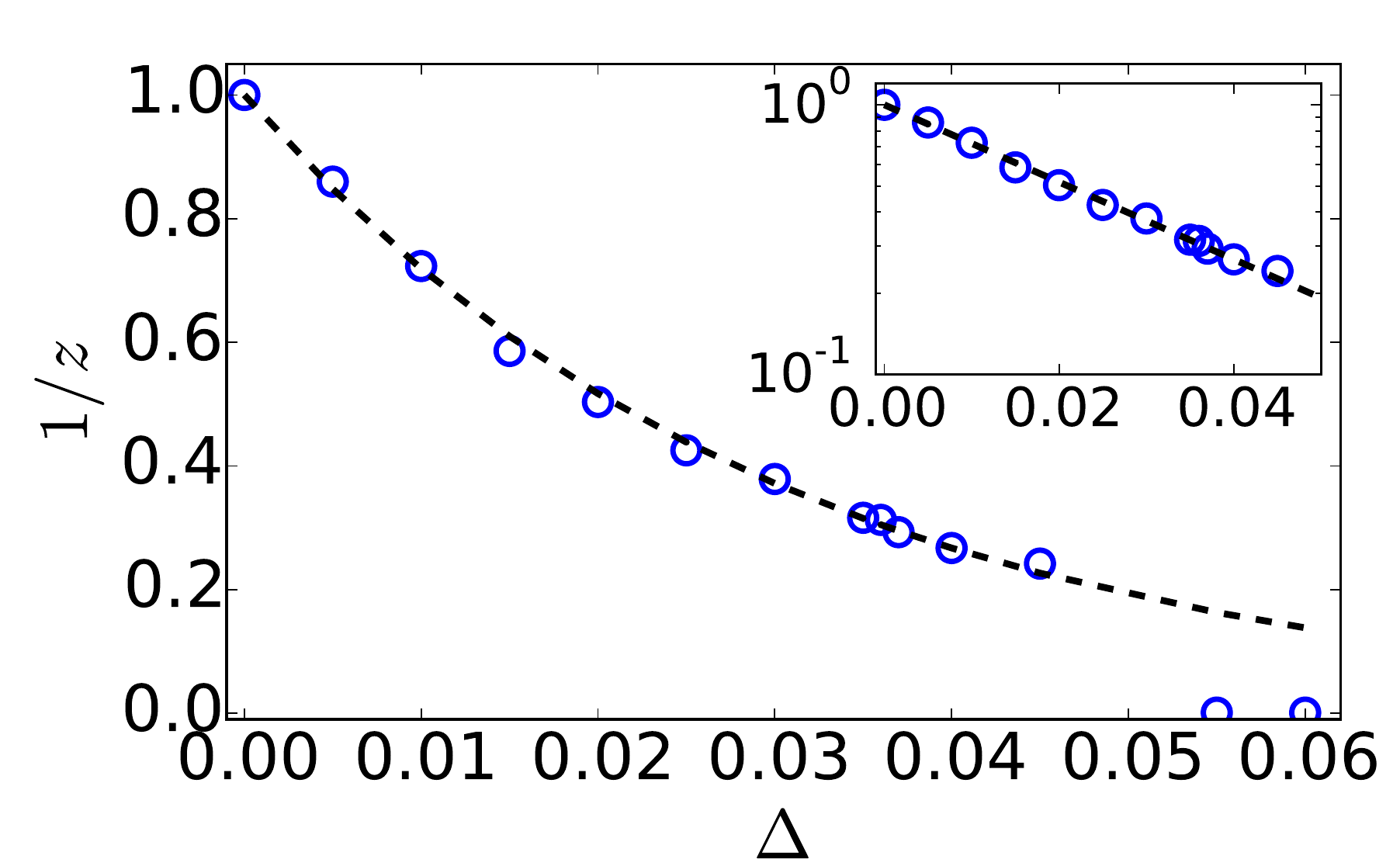}
\caption{Inverse dynamical exponent for the growth of the half-system entanglement entropy, plotted as a function of disorder.  Inset: same plot in lin-log scale, showing an exponential decrease of the inverse exponent at small disorder.}
\label{f.z}
\end{figure}
\end{center}

Fig.~\ref{f.z} shows the dynamical exponent $z$ as a function of disorder, extracted from power-law fits of post-quench growth of the half-system entanglement entropy (Fig.~2(b) of the main text). We observe a continuous disorder dependence of the $z$ exponent, with a seeming exponential increase with disorder, $z \approx \exp(\Delta/\Delta_0)$ for sufficiently weak disorder. For $\Delta \approx 5*10^{-2}$ it becomes difficult to distinguish between a power-law growth of the entropy with a small exponent and a logarithmic growth, and therefore we omit points in that range; on the other hand, for $\Delta = 6*10^{-2}$ our data are clearly consistent with a logarithmic growth over a sizable time range (see Fig.~2(c) of the main text), and therefore with a $z=\infty$ exponent.

\section{Pre-thermalization dynamics of the correlation functions after a quench}

\begin{center}
\begin{figure}[ht!]
\includegraphics[width=0.8\columnwidth]{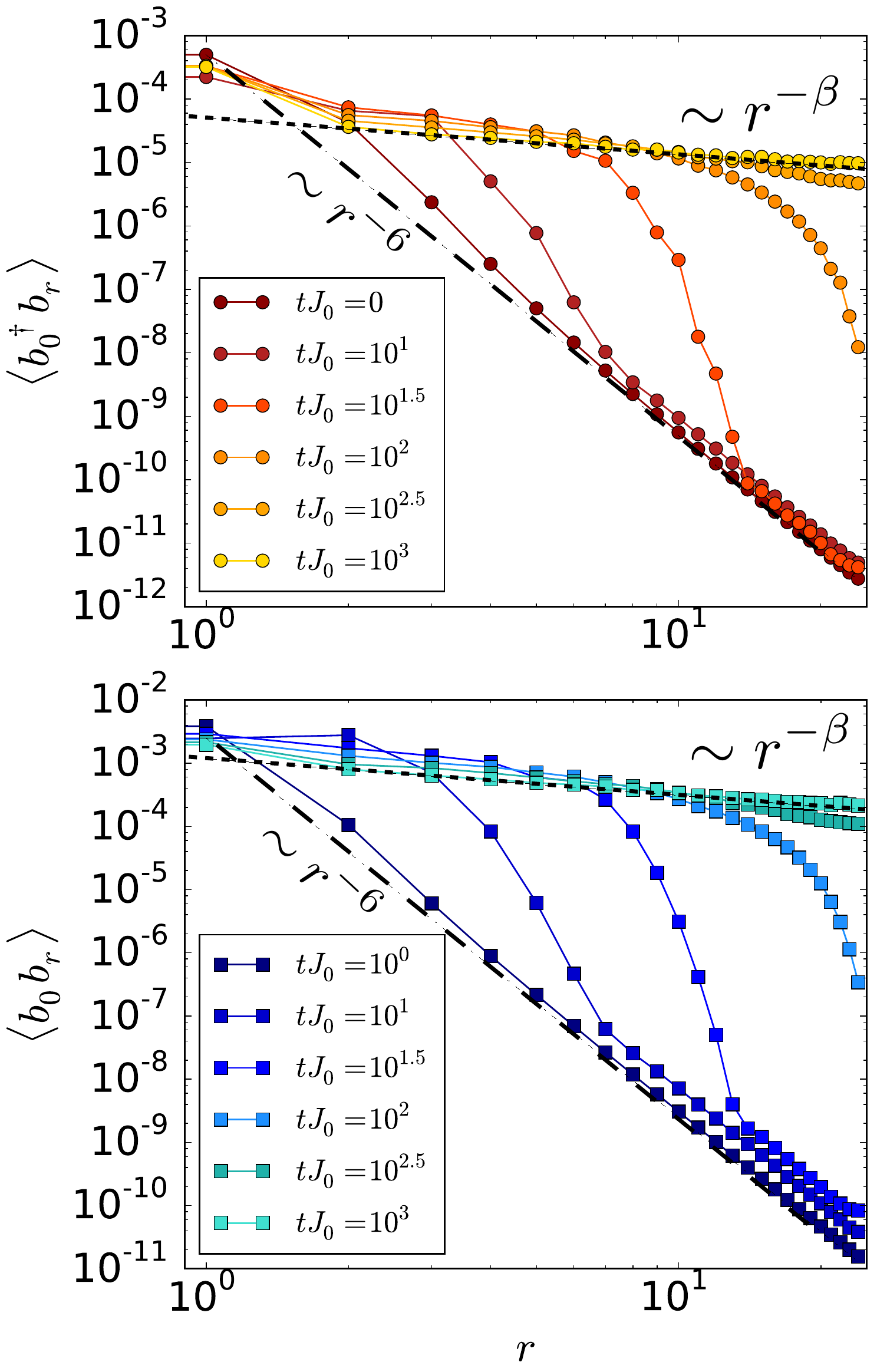}
\caption{Time evolution of the regular (upper panel) and anomalous (lower panel) bosonic correlation functions. One can observe the crossover from an initial decay as $r^{-6}$ to an asymptotic decay as $r^{-\beta}$ with $\beta \approx 0.58$. Here $\Delta = 2*10^{-2}$ and $L=50$.}
\label{f.asym}
\end{figure}
\end{center}

Fig.~\ref{f.asym} shows the pre-thermalization dynamics of the two-point correlation functions in the regime of anomalous diffusion ($\Delta = 2*10^{-2}$). At intermediate times we observe a progressive buildup of correlations at short distances, plus the immediate appearance of a $r^{-6}$ tail at long distances, due to the long-range nature of the vdW interactions. At long times, instead, correlations stabilize to a pre-thermalized form exhibiting a power-law decay, as $r^{-\beta}$, with $\beta \approx 0.58$. This form is clearly at odds with the form that one would generically expect at thermal equilibrium in this model (namely a long-distance decay as $r^{-6}$ stabilized by the long-range interactions). A discrepancy between the pre-thermalized state stabilized by the dephasing between the linear eigenmodes and the thermal equilibrium state is not surprising; yet the emergence of a power-law decay with a new exponent $\beta$ is a striking manifestation of the non-ergodic nature of the eigenmodes. 

\bibliography{biblio_disRyd}

\end{document}